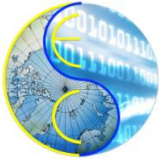

# Technical Report for HW2VEC – A Graph Learning Tool for Automating Hardware Security

Yasamin Moghaddas,

Tommy Nguyen,

Shih-Yuan Yu,

Rozhin Yasaei,

Mohammad Abdullah Al Faruque

Center for Embedded and Cyber-Physical Systems

University of California, Irvine

Irvine, CA 92697-2620, USA

{ymoghadd, tommytn1, shihyuay, ryasaei, alfaruqu}@uci.edu

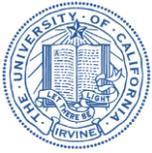

Department of Electrical Engineering and Computer Science,

University of California, Irvine

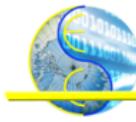

**CECS**

**CENTER FOR EMBEDDED & CYBER-PHYSICAL SYSTEMS**

## *Contents*

# 1 Introduction

The rapid growth in the hardware design complexity and the time-to-market pressure in combination have contributed to the globalization of the ***Integrated Circuit*** (IC) supply chain [1]. The lowered entry barrier of the IC has led to its spread across the globe, which has resulted in an increase in security threats [2]. For a ***System-on-Chip*** (SoC) company, any stage of the IC supply chain requires a vast investment of money and effort. For example, it costs $5 billion to develop a new foundry [3]. Thus, IC designers resort to ***Third-Party Electronic Automation*** (3P-EDA) tools and ***Third-Party Intellectual Property*** (3PIP) cores for cost reduction. However, such globalization also exposes the IC supply chain to hardware security threats such as ***Hardware Trojan Insertion***, ***IP Theft***, ***Overbuilding***, ***Counterfeiting***, ***Reverse Engineering***, and ***Covert & Side-Channel Attacks***.

In the literature [4], countermeasures and tools have been proposed to mitigate, prevent, or detect these security threats. For example, hardware-based primitives such as ***Physical Unclonable Functions*** (PUFs), ***True Random Number Generators*** (TRNGs), and cryptographic hardware can all intrinsically enhance security. Besides, the design tools built with countermeasures are critical in securing hardware in the early design phases. [5] and [6] have both contributed to HT detection with the former using a Hierarchical Temporal Memory (HTM) approach and the latter implementing neural networks. The existing works [7] [8] have successfully leveraged ***Machine Learning*** (ML) for detecting ***Hardware Trojans*** (HT) from hardware designs in both the ***Register Transfer Level*** (RTL) and ***Gate-Level Netlist*** (GLN) levels. However, to achieve the desired performance of enhancing security, these ML-based methods require a robust feature representation for a circuit in a ***non-Euclidean*** form that is challenging to acquire compared to finding the one from ***Euclidean*** data. Indeed, many basic objects like netlists or layouts are natural graph representations that conventional ML methods can hardly apply operations with (e.g., convolution on images). In the EDA field, existing works have utilized ***Graph Learning*** for test point insertion or fast power estimation in pre-silicon simulation. However, only a few approaches used graph learning for securing hardware during IC design phases due to the lack of supporting tools [9] [10].

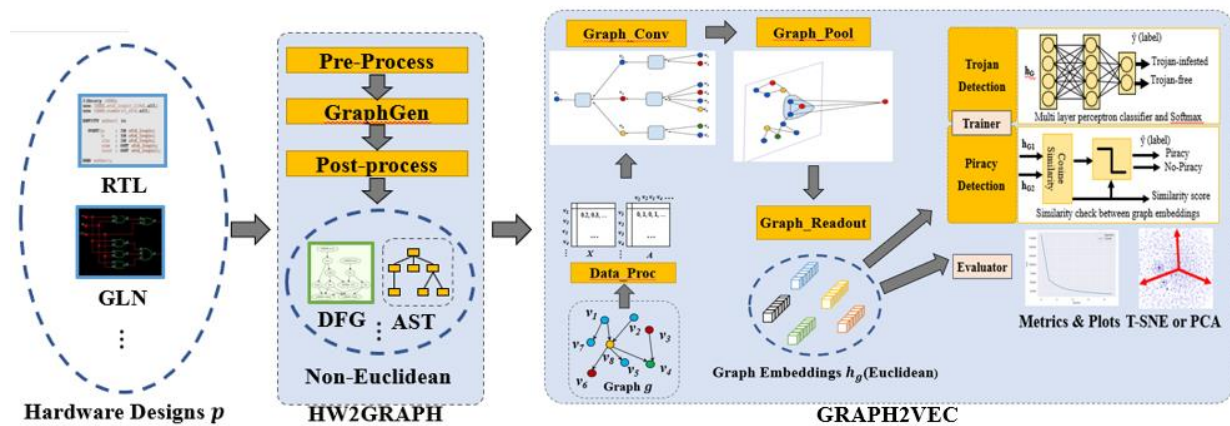

***Figure 1. The Overall Architecture of HW2VEC***

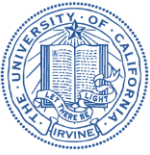 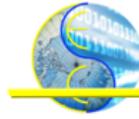

Department of Electrical Engineering and Computer Science,

University of California, Irvine

**CECS**

**CENTER FOR EMBEDDED & CYBER-PHYSICAL SYSTEMS**

In this technical report, we present ***HW2VEC*** [11], an open-source graph learning tool for hardware security, and its implementation details (Figure 1). ***HW2VEC*** provides toolboxes for graph representation extraction in the form of ***Data Flow Graphs*** (DFGs) or ***Abstract Syntax Trees*** (ASTs) from hardware designs at RTL and GLN levels. Besides, ***HW2VEC*** also offers graph learning tools for representing hardware designs in vectors that preserve both structural features and behavioral features. To the best of our knowledge, ***HW2VEC*** is the first open-source research tool that supports applying graph learning methods to hardware designs in different abstraction levels for hardware security. We organize the remainder of this technical report as follows: Section 2 introduces the architecture of ***HW2VEC***; Section 3 gives information about the use-case implementations; Section 4 provides the experimental results and demonstrates the performance of ***HW2VEC*** for two hardware security applications: HT detection and IP piracy detection; finally, Section 5 will conclude this report.

## 2    HW2VEC Architecture

As Figure 1 shows, our tool, ***HW2VEC***, consists of ***HW2GRAPH*** and ***GRAPH2VEC*** modules, with which a Euclidean representation of the hardware code is represented and used for learning. The responsibility of ***HW2GRAPH*** is to convert a hardware code ***p*** into a graph representation, ***g***, in the form of an AST or DFG. In hw2graph.py it carries out three stages: preprocessing (HW2GRAPH.preprocess), graph generation (HW2GRAPH.process), and postprocessing (using the process method in the respective graph class). ***GRAPH2VEC*** then creates a graph embedding model in models.py by executing the following steps: graph convolution (GRAPH_CONV), graph pooling (GRAPH_POOL), and graph readout (GRAPH_READOUT). In addition, ***GRAPH2VEC*** has a trainers.py module that allows the user to train the model so that it can detect HTs or IP piracy. The trainers.py module also allows the user to evaluate the model in order to discern its capability in detecting HTs and/or IP piracy. The user will also be able to visualize the vectorized graph and what the model has learned.

### 2.1 HW2GRAPH - Preprocessing Hardware Code

In `hw2graph.py`, there is the "***HW2GRAPH***" class, which has methods that flatten, remove comments and underscores from, and rename the top module of the hardware code. Modifications to the hardware code are critical because extracting the graph from the hardware code without making the necessary changes to it will result in issues in other stages of the HW2VEC down the line.

```
1.  def flatten(self, input_path, flattened_hw_path):
2.      flatten_content = ""
3.      all_containing_files = [Path(x).name for x in glob(fr'{input_path}/*.v'
4.  , recursive=True)]
5.      if "topModule.v" in all_containing_files:
6.          return
7.      for verilog_file in glob(fr'{input_path}/*.v'):
8.          with open(verilog_file, "r") as infile:
9.              flatten_content += infile.read()
10.     with open(flattened_hw_path, "w") as outfile:
11.         outfile.write(flatten_content)
```
***Listing 1. Preprocessing - Flattening of code***

Since a hardware design can contain several modules stored in separate files, the first step is to combine them into a single file through the process of flattening (Listing 1). The *flatten* method uses the *glob* function from the glob module in Python in order to find all files containing a specific *input_path*. For each file that is found, we append its contents to a string, called *flatten_content*, which is initially empty (line 7). Once we traverse through all the files, we write the *flatten_content* string to an output file, called *outfile*. Ultimately, this process ends up taking all the contents of files with a specific *input_path* and putting them into a single file.

```
1.   modules_dic={}
2.   for line in lines:
3.       words = line.split()
4.       for word_idx, word in enumerate(words):
5.           if word == 'module':
6.               module_name = words[word_idx+1]
7.               if '(' in module_name:
8.                   idx = module_name.find('(')
9.                   module_name = module_name[:idx]
10.                  modules_dic[module_name]= 1
11.
12.              else:
13.                  modules_dic[module_name]= 0
```
***Listing 2. Preprocessing - Renaming Top Module***

Once we remove comments and underscores from the flattened code, we rename the top module. We must start at the top in order to detect HT at the earliest point; therefore, it is important to identify the top module. *HW2GRAPH.rename_topModule* initializes an empty dictionary, called *modules_dic*, which is meant to store module names as keys and the module name frequency as the values.

We then traverse through the lines in *hw_path* and add any module name that is prefaced by the word *module* to the dictionary, initializing the value to 1 if there is a parenthesis in the module name and a 0 otherwise (Listing 2).

```
1.   for line in lines:
2.       words = line.split()
3.       for word in words:
4.           if word in modules_dic.keys():
5.               modules_dic[word] += 1
6.
7.   for m in modules_dic:
8.       if modules_dic[m] == 1:
9.           top_module = m
10.          break
```
***Listing 3. Preprocessing - Renaming Top Module***

As Listing 3 shows, we traverse through the lines again and if a word is a module name that is in the dictionary, we increment its frequency by 1 (line 5). As can be seen on line 8, the module name that has a frequency of 1 is the top module, and for each line that contains the *top_module* name as a word, we replace it with the word "top". Once this step has finished, we return the *flattened_hw_path* and then begin processing (described in the next section).

## 2.2 HW2GRAPH - Graph Generation from Hardware Code

After the preprocessing has completed, the hardware code must be converted into a graph representation: a DFG or an AST. In a DFG, the nodes usually represent mathematical operations and the edges represent the input and output dependencies between the nodes [12].Figure 2 is a graphical representation of a DFG. The variables $a$, $b$, $c$, and $d$ represent the inputs; the circles represent the nodes, which are the mathematical operations (multiplication, addition, and division); $X$, $Y$, and $Z$ are the outputs. Their responsibility in **HW2VEC** is to indicate the relationships and dependencies between the circuit's signals and give a higher-level expression of the code's computational structure.

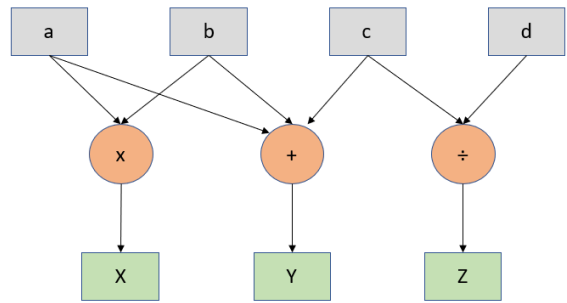

*Figure 2. An example of a DFG*

An AST is a tree representation of the code, where each node is associated with a type that it represents [13]. More specifically, the nodes could represent keywords, variables, etc. and the edges show the relationship between them. Figure 3 gives an example of an AST of Verilog code that takes a variable, *counter*, and increments it by 1 as long as *counter* is less than or equal to 10. The purpose of ASTs in **HW2VEC** and in general is to capture the syntactic structure of hardware code.

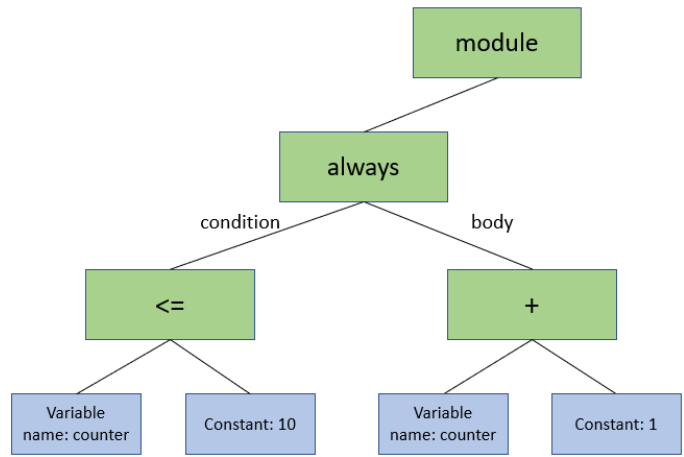

*Figure 3. An example of an AST for Verilog code*

In order to carry out graph generation, we integrate **PyVerilog**, a hardware design toolkit for parsing the Verilog code, into this module [14]. The code is first converted into a parse tree with the help of **YACC** (Yet Another Compiler-Compiler), a lexical analyzer that gives a structural representation of the hardware code [15].

```python
1.   def _generate_ast_dict(self, ast_node):
2.       class_name = ast_node.__class__.__name__
3.       structure = {}
4.       #based on the token class_name, determine the value type of class_name
5.       if class_name in self.ARRAY_GEN:
6.           structure[class_name] = [getattr(ast_node, n) for n in ast_node.attr_names] if
     ast_node.attr_names else []
7.           for c in ast_node.children():
8.               structure[class_name].append(self._generate_ast_dict(c))
9.       elif class_name in self.DICTIONARY_GEN:
10.          structure[class_name] = self._generate_ast_dict(ast_node.children()[0])
11.      elif class_name in self.CONST_DICTIONARY_GEN:
12.          structure = {}
13.          structure[class_name] = getattr(ast_node,ast_node.attr_names[0])
14.          return structure
15.      else:
16.          raise Exception(f"Error. Token name {class_name} is invalid or has not yet been
     supported")
17.      return structure
```

***Listing 4. How AST is generated in JSON format***

In order to produce the AST, we use *ASTGenerator._generate_ast_dict* method (Listing 4). This method iterates recursively through each node of the parse tree using a Depth First Search (DFS). At each recursive step, we determine whether to construct a collection of name/value pairs, an ordered list of values, or a single name/value pair based on the token names used in Verilog AST; this can be seen by the if/elif statements on lines 5, 9, 11, and 15. It is important to note that the AST is in **JSON** (JavaScript Object Notation) format, which is a lightweight data format that is known for being readable and easy to parse by machines.

In order to produce a DFG, *DFGGenerator.process* creates an instance of the *VerilogDataflowAnalyzer* class, called *dataflow_analyzer*. The *VerilogDataflowAnalyzer.generate* outputs a parse tree also with the help of YACC. An instance of the *VerilogGraphGenerator* is then created and assigned to the variable name *dfg_graph_generator*. The *dfg_graph_generator* has a *binddict*, which is a dictionary whose keys are nodes (the signals) and the values are each node's associated dataflow object. For each signal in the *dfg_graph_generator binddict*, we call the *DFGGenerator.generate* to create a signal DFG. Lastly, we merge all the signal DFGs together. The resulting graph is a DFG that is also in JSON format.

The resulting graph, either DFG or AST, is denoted as $g = (V, E)$, where $V$ is a set of nodes which can be operators and $E$ is a set of edges, which represent the relation between the nodes. The AST is a tree type of graph in which the nodes $V$ can be operators (mathematical, gates, loop, conditional, etc.), signals, or attributes of signals. The edges $E$ indicate the relation between nodes. The DFG shows data dependency where each node in $V$ represents signals, constant values, and operations such as xor, and, concatenation, branch, or branch condition, etc. Each edge in $E$ stands for the data dependency relation between two nodes. Specifically, for all $v_i$, $v_j$ pairs, the edge $e_{ij}$ belongs to $E$ ($e_{ij} \in E$) if $v_i$ depends on $v_j$, or if $v_j$ is applied on $v_i$.

## 2.3 HW2GRAPH - Post-Processing

After processing, the graph *g* will be in JSON format. In the post-processing stage, we convert a JSON-formatted graph into a NetworkX graph object. NetworkX is a Python package that allows for the creation and manipulation of graphs and networks. It is an efficient, scalable, and highly portable framework for graph analysis. In later stages we will be making use of the PyTorch-Geometric library, a geometric deep learning extension for PyTorch. PyTorch is a machine learning library for Python programs [16]. Libraries like PyTorch Geometric take a NetworkX graph object as their primary data structure in their pipelines, so a conversion from JSON format to a NetworkX format is necessary.

```
1.  def process(self, verilog_file):
2.          #when generating AST, determines which substructure (dictionary/array) to generate
3.      #before converting the json-like structure into actual json
4.
5.          self.ast, _ = parse([verilog_file], debug=False)
6.          ast_dict = self._generate_ast_dict(self.ast)
7.
8.          nx_graph = nx.DiGraph()
9.          for key in ast_dict.keys():
10.             self._add_node(nx_graph, 'None', key, ast_dict[key])
11.
12.         return nx_graph
```

*Listing 5. ASTGenerator.process - AST in JSON format converted to a NetworkX graph object*

For both the DFG and AST, we initialize the NetworkX graph object, which we name *nx_graph*, as a NetworkX DiGraph object, which holds directed edges. In *ASTGenerator.process*, we use the *ASTGenerator._add_node* on line 10 (Listing 5) to add nodes and edges to the NetworkX graph object based on the keys and values in the *ast_dict*.

```
1.  nx_graph = nx.DiGraph()
2.
3.  for node in dfg_graph_generator.graph.nodes():
4.      node_name = node.name
5.      if '_graphrename' in node.name:
6.          node_name = node.name[:node.name.index('_graphrename')]
7.      if '.' in node_name:
8.          type_of_node = node_name.split('.')[-1]
9.      elif '_' in node_name:
10.         type_of_node = node_name.split('_')[-1]
11.     else:
12.         type_of_node = node_name.lower()
13.     nx_graph.add_node(node.name, label=type_of_node)
14.     for child in dfg_graph_generator.graph.successors(node):
15.         nx_graph.add_edge(node.name, child.name)
16.
17. return nx_graph
```

*Listing 6. DFGGenerator.process - DFG in JSON format converted to a NetworkX graph object*

In *DFG.process*, we iterate through the nodes of the *dfg_graph_generator.graph* (a DFG graph object in JSON format) and use the *add_node* method of a NetworkX graph to add nodes to the DiGraph object, making sure to specify the appropriate node name and node type (Listing 6).

## 2.4 GRAPH2VEC - Dataset Processor

The *DataProcessor* class then normalizes the NetworkX graph object by iterating through the nodes of the *nx_graph* and giving each node a label that indicates its type. For the DFG, the type of the node can be *numeric*, *output*, *input*, or *signal*. The AST node can have a type of *names* or *pure numeric*, otherwise the type remains unchanged. The label type is then used to convert each one of the nodes into a vectorized representation.

Essentially, the *DataProcessor.process* (Listing 7) converts the graph *g* into the matrices **X**, which represents the node embeddings, and **A**, which represents the adjacency information of the graph. This happens in *DataProcessor.process* when we call *from_networkx*. The *from_networkx* method on line 3 of Listing 7 is imported from *torch_geometric.utils.convert*, which is a module in pytorch_geometric. *from_networkx* has an empty dictionary, *data*, which is filled with the graph's nodes and edges. By calling *torch.tensor* on each item in *data*, we create feature vectors for each key in the dictionary. These feature vectors represent the node embeddings, **X**. *from_networkx* also creates an adjacency matrix (this represents **A**) of the NetworkX graph object's edges by calling *torch.LongTensor* on the list of the graph's edges. This adjacency matrix, called *edge_index*, becomes an attribute of a dictionary, called *data*. Finally, *data* is converted into a PyTorch geometric Data instance by calling *torch_geometric.data.Data.from_dict* on the dictionary and this object is returned by the function.

```
1.  def process(self, nx_graph):
2.      self.normalize(nx_graph)
3.      data = from_networkx(nx_graph)
4.      data.hw_name = nx_graph.name
5.      data.hw_type = nx_graph.type
6.      self.graph_data.append(data)
```
*Listing 7. DataProcessor normalizes graph and creates matrices X and A*

## 2.5 GRAPH2VEC - Graph Embedding Model: Graph Convolution

In literature, Graph Convolution has a message propagation phase, which involves two sub-functions: **AGGREGATE** and **COMBINE** functions. The **AGGREGATE** function updates the node embeddings after each $k$-th iteration to produce $\mathbf{X}^{(k)}$ using each node representation $h_v^{(k-1)}$ in $\mathbf{X}^{(k-1)}$. The function essentially accumulates the features of the neighboring nodes and produces an aggregated feature vector $a_v^{(k)}$ for each layer k. The **COMBINE** function combines the previous node feature $h_v^{(k-1)}$ with $a_v^{(k)}$ to output the next feature vector $h_v^{(k)}$. The final node embedding after the message propagation is denoted as $\mathbf{X}^{prop}$.

```
1.  class GRAPH_CONV(nn.Module):
2.      def __init__(self, type, in_channels, out_channels):
3.          super(GRAPH_CONV, self).__init__()
4.          self.type = type
5.          self.in_channels = in_channels
6.          if type == "gcn":
7.              self.graph_conv = GCNConv(in_channels, out_channels)
8.
9.      def forward(self, x, edge_index):
10.         return self.graph_conv(x, edge_index)
```
*Listing 8. models.GRAPH_CONV*

Our tool achieves the convolution operation described above through the *GRAPH_CONV* class (Listing 8). The *GRAPH_CONV* class performs graph convolutions by first creating a *GCNConv* object. The *GCNConv* class, which is imported from *torch_geometric.nn. GCNConv*, is initialized by a graph convolution *type* (string), *in_channels* (int), which is the size of the input (the number of nodes), and *out_channels* (int), which is the size of the output. The forward function has parameters *x* and *edge_index*, which are basically **X** and **A**. The forward function on line 9 returns the result of calling the *propagate* method, which is inherited from the *MessagePassing* class in torch_geometric.nn.conv. The *propagate* function internally calls the *message*, *aggregate*, and *update* methods. The *message* function normalizes the neighboring node features; the *aggregate* method does the same job as **AGGREGATE** and **COMBINE**. So, the forward function produces an updated feature matrix, $\mathbf{X}^{prop}$, which is returned by the *update* method. To track the convolution layers, there is *GRAPH2VEC.set_graph_conv*, which basically creates a *layers* list of the convolutions (Listing 9).

```
1.  def set_graph_conv(self, convs):
2.      self.layers = []
3.
4.      for conv in convs:
5.          conv.to(self.config.device)
6.          self.layers.append(conv)
7.      self.layers = nn.ModuleList(self.layers)
```
*Listing 9. models.GRAPH2VEC.set_graph_conv*

## 2.6 GRAPH2VEC - Graph Embedding Model: Graph Pooling

The following stage is graph pooling, which uses an attention-based pooling layer to target a specific part of the graph. In this layer, a *top-k filtering* is performed on the nodes according to the scoring results:

$$\alpha = \text{SCORE}(\mathbf{X}^{prop}, \mathbf{A})$$

$$\mathbf{P} = \text{top}_k(\alpha)$$

where α stands for the coefficients predicted by the graph pooling layer for the nodes. **P** is the indices of the pooled nodes, which are chosen from the top *k* of the nodes ranked according to α. The number *k* used in top-k filtering is calculated by a pre-defined pooling ratio, pr using $k = pr \times |V|$, where we consider only a constant fraction *pr* of the embeddings of the nodes of the DFG to be relevant (i.e., 0.5). The node embeddings and edge adjacency information after pooling are the following:

$$\mathbf{X}^{pool} = (\mathbf{X}^{prop} \odot \tanh(\alpha))\mathbf{P}$$

$$\mathbf{A}^{pool} = \mathbf{A}^{prop}(\mathbf{P}, \mathbf{P})$$

where (**P**,**P**) is the information of the adjacency matrix between the nodes in the subset.

```
1.  class GRAPH_POOL(nn.Module):
2.      def __init__(self, type, in_channels, poolratio):
3.          super(GRAPH_POOL, self).__init__()
4.          self.type = type
5.          self.in_channels = in_channels
6.          self.poolratio = poolratio
7.          if self.type == "sagpool":
```

```
8.                self.graph_pool = SAGPooling(in_channels, ratio=poolratio)
9.          elif self.type == "topkpool":
10.             self.graph_pool = TopKPooling(in_channels, ratio=poolratio)
11.
12.     def forward(self, x, edge_index, batch):
13.         return self.graph_pool(x, edge_index, batch=batch)
```
*Listing 10.  models.GRAPH_POOL*

Our tool performs graph pooling through the *GRAPH_POOL* class in models.py (Listing 10). In order to initialize a *GRAPH_POOL* instance, the user must pass in a type (string: *topkpool* or *sagpool*), the size of the input (int value assigned to *in_channels* parameter), and the pooling ratio, *poolratio*. Depending on the type that is passed to the constructor, we either create an instance of the *SAGPooling* class (self-attention graph pooling) or the *TopKPooling* class, both of which have been imported from torch_geometric.nn. Both *SAGPooling* and *TopKPooling* are instantiated by the same values assigned to *in_channels* and *poolratio* as the *GRAPH_POOL* class. *SAGPooling.forward* carries out the self-attention graph pooling and takes feature vectors *x*, an adjacency matrix *edge_index*, and *batch* (a vector that indicates which nodes in the batch are a part of the same graph) as parameters [17]. Essentially, the forward function returns $\mathbf{X}^{pool}$ and $\mathbf{A}^{pool}$ by performing topk-filtering, using the pooling ratio that was passed as an argument to the *SAGPool* constructor.

*TopKPooling* is done in the same manner as *SAGPooling* (forward method is identical). The only difference between the two classes is their *reset_parameters* method and the fact that *TopKPooling* has a weight attribute.

## 2.7 GRAPH2VEC - Graph Embedding Model: Graph Readout

Finally, the *GRAPH_READOUT* (Listing 11) will sum up or average up the node features $\mathbf{X}^{pool}$ to produce the node embeddings $h_g{}^{(k)}$ for each graph, *g*. In order to create an instance of *GRAPH_READOUT* the user only needs to pass a *type* (string) to the constructor.

```
1.  class GRAPH_READOUT(nn.Module):
2.      def __init__(self, type):
3.          super(GRAPH_READOUT, self).__init__()
4.          self.type = type
5.
6.      def forward(self, x, batch):
7.          if self.type == "max":
8.              return global_max_pool(x, batch)
9.          elif self.type == "mean":
10.             return global_mean_pool(x, batch)
11.         elif self.type == "add":
12.             return global_add_pool(x, batch)
```
*Listing 11.  models.GRAPH_READOUT*

Depending on the type, the forward function calls *global_max_pool*, *global_mean_pool*, or *global_add_pool*, as indicated by the if/elif statements on lines 7, 9, and 11 of Listing 11. The function *global_add_pool* takes *x*, a node feature matrix (to which $\mathbf{X}^{pool}$ is assigned to), and *batch*, a vector that indicates which nodes in the batch are a part of the same graph. The output of *global_add_pool* for a single graph $G_i$ is calculated by the following formula:

$$r_i = \sum_{n=1}^{N_i} x_n$$

where $i$ is the $i$-th level of the graph, $N_i$ is the $i$-th node feature, and $x_n$ is a node feature vector. Basically, *global_add_pool* returns batch-wise graph-level-outputs by adding node features across the node dimension for the graph $G_i$ [18].

For computing the average, the function *global_mean_pool* uses the following formula to return batch-wise graph-level-outputs by averaging node features across the node dimension for the graph $G_i$ [18]:

$$r_i = \frac{1}{N_i} \sum_{n=1}^{N_i} x_n$$

The user will be able to use the resulting graph embedding $r_i$ (same as $h_g^{(k)}$) to model the behavior of circuits (use $h_g$ for simplicity). After the graph readout step, the fixed-length embeddings of hardware designs then become compatible with ML algorithms.

## 2.8 GRAPH2VEC - Trainer and Evaluator

The trainers.py module takes training datasets, validating datasets, and a set of hyperparameter configurations to train a GNN model. HW2VEC currently supports two types of trainers, a *GraphTrainer* class and *PairwiseGraphTrainer* class. *GraphTrainer* and *PairwiseGraphTrainer* both inherit from the *BaseTrainer*. In its *build* method, *BaseTrainer* creates an instance of an Adam optimizer, which implements the Adam algorithm. There is also a *get_embeddings* method that returns a list of the graph embeddings and *BaseTrainer.visualize_embeddings*, which allows the user to visualize the vectorized graph. *BaseTrainer.visualize_embeddings* can be called by passing a *DataLoader* object and a path. The *DataLoader* class joins data objects (in this case, a list of NetworkX graph objects) from a *torch_geometric.data.dataset* to a mini-batch.

```
1.  def train(self, train_loader, test_loader):
2.      tqdm_bar = tqdm(range(self.config.epochs))
3.
4.      for epoch_idx in tqdm_bar:
5.          self.model.train()
6.          acc_loss_train = 0
7.
8.          for data in train_loader:
9.              self.optimizer.zero_grad()
10.             graph1, graph2, labels = data[0].to(self.config.device),
        data[1].to(self.config.device), data[2].to(self.config.device)
11.
12.             loss_train = self.train_epoch_ip(graph1, graph2, labels)
13.             loss_train.backward()
14.             self.optimizer.step()
15.
16.             acc_loss_train += loss_train.detach().cpu().numpy()
17.
```

```
18.              tqdm_bar.set_description('Epoch: {:04d}, loss_train: {:.4f}'.format(epoch_idx,
         acc_loss_train))
19.
20.            if epoch_idx % self.config.test_step == 0:
21.                self.evaluate(epoch_idx, train_loader, test_loader)
```

*Listing 12. trainers.PairwiseGraphTrainer.train*

The *PairwiseGraphTrainer* class, which inherits methods from *BaseTrainer*, considers pairs of graphs, calculates their similarities, and performs the graph similarity learning and evaluation. Users can use this class for IP piracy detection by determining whether one of the two hardware designs is stolen from the other or not. To implement [9], the GNN model must be trained with a graph-pair classification trainer in *GRAPH2VEC*. The user must pass a *DataLoader* object of the train graphs and a *DataLoader* object of the test graphs to *PairwiseGraphTrainer.train* in order to call the function so that the GNN model is trained (Listing 12). The paragraph that follows describes the methods that are called each time we iterate through all the graph objects in the *train_loader* in order to carry out the process of training the model.

```
1.   def train_epoch_ip(self, graph1, graph2, labels):
2.       g_emb_1, _ = self.model.embed_graph(graph1.x, graph1.edge_index, batch=graph1.batch)
3.       g_emb_2, _ = self.model.embed_graph(graph2.x, graph2.edge_index, batch=graph2.batch)
4.
5.       g_emb_1 = self.model.mlp(g_emb_1)
6.       g_emb_2 = self.model.mlp(g_emb_2)
7.
8.       loss_train = self.cos_loss(g_emb_1, g_emb_2, labels)
9.       return loss_train
```

*Listing 13. trainers.PairwiseGraphTrainer.train_epoch_ip*

The first step is to use *HW2GRAPH* to convert a pair of circuit designs $p_1$, $p_2$ into a pair of graphs $g_1$, $g_2$. Then, *GRAPH2VEC* transforms both $g_1$ and $g_2$ into graph embeddings $h_{g1}$, $h_{g2}$. *PairwiseGraphTrainer.train_epoch_ip* takes two graphs and their labels as arguments and calls *model.embed_graph* to obtain $h_{g1}$ and $h_{g2}$ (Listing 13) . The user must pass the graph's feature vector attribute (*x*), adjacency matrix attribute (*edge_index*), and batch to *model.embed_graph* in order to call it (lines 2 and 3 of Listing 13). Then, the *cos_loss* function is called in *train_epoch_ip* to return the cosine embedding loss, which is a way of measuring whether two embeddings are similar or not using the cosine distance (Listing 13, line 8). This loss is added to the *total_loss* variable.

```
1.   def evaluate(self, epoch_idx, train_loader, test_loader):
2.       train_loss, train_labels, _, train_preds = self.inference(train_loader)
3.       test_loss, test_labels, _, test_preds = self.inference(test_loader)
4.
5.       print("")
6.       print("Mini Test for Epochs %d:"%epoch_idx)
7.
8.       self.metric_calc(train_loss, train_labels, train_preds, header="train")
9.       self.metric_calc(test_loss,  test_labels,  test_preds,  header="test ")
10.
11.      if self.min_test_loss >= test_loss:
12.          self.model.save_model(str(self.config.model_path_obj/"model.cfg"),
         str(self.config.model_path_obj/"model.pth"))
13.
14.          # on final evaluate call
```

```
15.        if(epoch_idx==self.config.epochs):
16.            self.metric_print(self.min_test_loss, **self.metrics, header="best ")
```

***Listing 14. trainers.PairwiseGraphTrainer.evaluate***

At certain points, we call the evaluate method by passing the *epoch_idx*, the train loader, and the test loader as arguments (Listing 14). One lines 2 and 3, the function calls the *inference* method on both the train loader and the test loader. For each dataset in the *DataLoader* object that was passed to *inference*, *inference* calls *inference_epoch_ip* by passing the *graph1* and *graph2* (both are NetworkX graph objects) to *inference_epoch_ip*.

```
1.  def inference_epoch_ip(self, graph1, graph2):
2.      g_emb_1, _ = self.model.embed_graph(graph1.x, graph1.edge_index, batch=graph1.batch)
3.      g_emb_2, _ = self.model.embed_graph(graph2.x, graph2.edge_index, batch=graph2.batch)
4.
5.      g_emb_1 = self.model.mlp(g_emb_1)
6.      g_emb_2 = self.model.mlp(g_emb_2)
7.
8.      similarity = self.cos_sim(g_emb_1, g_emb_2)
9.      return g_emb_1, g_emb_2, similarity
```

***Listing 15. trainers.PairwiseGraphTrainer.inference_epoch_ip***

*inference_epoch_ip* computes the graph embeddings for each graph to yield *g_emb_1* and *g_emb2*, as shown on lines 2 and 3 of Listing 15. On line 8 we call *cos_sim(g_emb_1, g_emb2)* in order to calculate the cosine similarity of the two graph embeddings. With each iteration, we append the similarity to the *outputs* list. The goal of the *inference* function is to return the average loss, a matrix of the labels, a matrix of *outputs* (cosine similarities), and a matrix of the predictions for IP piracy detection based on cosine similarities. These three values are then passed to the *metric_calc* method, which calculates the accuracy score (the correctly predicted ratio) , F1 score (the weighted average of precision), precision score, and recall score. The precision score is calculated as the number of true positives divided by the sum of true positives and false positives. The recall score is the number of true positives divided by the sum of true positives and false negatives. These values are all stored in a dictionary called *matrix*, where the keys are strings that denote the type of score (e.g. *acc* for accuracy score) and the values are the corresponding computed scores. *metric_calc* then calls *metric_print* which prints the loss, accuracy score, confusion matrix, precision score, recall score, and the header. In summary, the *evaluate* method prints out scores that allow the user to assess the degree to which the model has learned from training.

```
1.  def train(self, data_loader, valid_data_loader):
2.      tqdm_bar = tqdm(range(self.config.epochs))
3.
4.      for epoch_idx in tqdm_bar:
5.          self.model.train()
6.          acc_loss_train = 0
7.
8.          for data in data_loader:
9.              self.optimizer.zero_grad()
10.             data.to(self.config.device)
11.
12.             loss_train = self.train_epoch_tj(data)
13.             loss_train.backward()
14.             self.optimizer.step()
```

```
15.             acc_loss_train += loss_train.detach().cpu().numpy()
16.
17.             tqdm_bar.set_description('Epoch: {:04d}, loss_train: {:.4f}'.format(epoch_idx,
    acc_loss_train))
18.
19.         if epoch_idx % self.config.test_step == 0:
20.             self.evaluate(epoch_idx, data_loader, valid_data_loader)
```

*Listing 16. trainers.GraphTrainer.train*

In order to train the model for HT detection, the user should use the *GraphTrainer* class. An instance of *GraphTrainer* is created by passing a graph and assigning the result of a call to *DataProcessor.get_class_weights* on the train graphs and assigning it to the *class_weights* parameter. The user must then call the trainer's *build* method on the model, and then call the *train* method (Listing 16) on the train loader and the valid loader, both of which are instances of *DataLoader* for the train graphs and test graphs, respectively.

```
1.  def train_epoch_tj(self, data):
2.      output, _ = self.model.embed_graph(data.x, data.edge_index, data.batch)
3.      output = self.model.mlp(output)
4.      output = F.log_softmax(output, dim=1)
5.
6.      loss_train = self.loss_func(output, data.label)
7.      return loss_train
```

*Listing 17. trainers.GraphTrainer.train_epoch_tj*

*GraphTrainer.train* passes each dataset in the train loader to *GraphTrainer.train_epoch_tj* (Listing 17), which transforms each graph, $g$, into a graph embedding $h_g$. Then it uses $h_g$ to make a prediction $\hat{y}$ with an MLP layer. Ultimately, it returns the cross-entropy loss of the graph (Listing 17, line 6), which is then added to the *acc_loss_train*, the accumulated loss. The collective cross-entropy loss $L$ of all the graphs in the training set is equivalent to the following equation:

$$L = H(Y, \hat{Y}) = \sum_i \quad y_i * log_e(\hat{y_i}),$$

where $H$ is the loss function. $Y$ stands for the set of ground-truth labels (either TROJAN or NON-TROJAN) and $\hat{Y}$ represents the corresponding set of predictions.

```
1.  def inference(self, data_loader):
2.      labels = []
3.      outputs = []
4.      node_attns = []
5.      total_loss = 0
6.      folder_names = []
7.
8.      with torch.no_grad():
9.          self.model.eval()
10.         for i, data in enumerate(data_loader):
11.             data.to(self.config.device)
12.
13.             loss, output, attn = self.inference_epoch_tj(data)
14.             total_loss += loss.detach().cpu().numpy()
15.
16.             outputs.append(output.cpu())
```

```
17.
18.                 if 'pool_score' in attn:
19.                     node_attn = {}
20.                     node_attn["original_batch"] = data.batch.detach().cpu().numpy().tolist()
21.                     node_attn["pool_perm"] = attn['pool_perm'].detach().cpu().numpy().tolist()
22.                     node_attn["pool_batch"] = attn['batch'].detach().cpu().numpy().tolist()
23.                     node_attn["pool_score"] = attn['pool_score'].detach().cpu().numpy().tolist()
24.                     node_attns.append(node_attn)
25.                 labels += np.split(data.label.cpu().numpy(), len(data.label.cpu().numpy()))
26.         outputs = torch.cat(outputs).reshape(-1,2).detach()
27.         avg_loss = total_loss / (len(data_loader))
28.
29.         labels_tensor = torch.LongTensor(labels).detach()
30.             outputs_tensor = torch.FloatTensor(outputs).detach()
31.             preds = outputs_tensor.max(1)[1].type_as(labels_tensor).detach()
32.
33.         return avg_loss, labels_tensor, outputs_tensor, preds, node_attns
```

***Listing 18. trainers.GraphTrainer.inference***

At certain points, *GraphTrainer.train* will call *GraphTrainer.evaluate* on the train loader and the valid loader, which in turn calls *GraphTrainer.inference* (Listing 18) on both the train loader and the valid loader separately. As the function iterates through train loader, it passes each dataset in the train loader to *GraphTrainer.inference_epoch_tj*, which does the same exact computations as *train_epoch_tj*, except that in addition to the loss, it returns the graph embedding, output (a matrix of the graph embedding after *GRAPH_READOUT* is performed), and the data object's attention weights in the form of a dictionary, called *attn*. The loss is added to the *total_loss* variable and the *output* is appended to a list, called *outputs*. If there is a pool score in *attn*, then we add it to a list, called *node_attns*. In the end the function ends up calculating the average loss, creating a matrix out of the outputs, labels, and predictions (which is computed based on the values in the outputs matrix, which holds the graph embeddings), and returns all the aforementioned data. Ultimately, *GraphTrainer.evaluate* ends up printing these values to allow the user to get a better understanding about the effectiveness of the training based on the model's ability to learn from the trainer.

## 3 Use-Case Implementations and Explanations

In this section, we describe the HW2VEC use cases. There are three use-cases that we provide. The first use-case exhibits a fundamental case in which hardware design *p* is converted into a graph *g* and then into a fixed-length embedding $h_g$. The second use-case demonstrates how HW2VEC can be applied for the task of hardware Trojan detection. The third use case shows how HW2VEC can be applied for the task of hardware IP piracy detection. We will demonstrate these three use-cases in the following three subsections.

### 3.1 Use-case 1: Converting a Hardware Design to a Graph Embedding

The first use-case demonstrates the transformation of a hardware design *p* into a graph *g* and then into an embedding $h_g$. HW2GRAPH uses *preprocessing* (PRE_PROC), *graph generation* (GRAPH_GEN), and *post-processing* (POST_PROC) modules which are detailed in Section 2 to convert each hardware design into the corresponding graph *g*. The *g* is fed to GRAPH2VEC with the use of the *Data Processing* (DATA_PROC) module to generate features $X$ and adjacency matrix $A$. Then, $X$ and $A$ are processed through *GRAPH_CONV*, *GRAPH_POOL*, and *GRAPH_READOUT* layers to generate the graph embedding $h_g$. The resulting $h_g$ can be further inspected with the utilities of the *Evaluator* module (see Section 2). The implementation of use-case 1 is provided in `use_case_1.py` of our repository and is also shown below as follows:

```
1.  def use_case_1(cfg, hw_design_dir_path, pretrained_model_weight_path, pretrained_model_cfg_path):
2.      hw2graph = HW2GRAPH(cfg)
3.
4.      hw_design_path = hw2graph.preprocess(hw_design_dir_path)
5.      hardware_nxgraph = hw2graph.process(hw_design_path)
6.
7.      data_proc = DataProcessor(cfg)
8.      data_proc.process(hardware_nxgraph)
9.      vis_loader = DataLoader(data_proc.get_graphs(), batch_size=1)
10.
11.     model = GRAPH2VEC(cfg)
12.     model.load_model(pretrained_model_cfg_path, pretrained_model_weight_path)
13.     model.to(cfg.device)
14.     graph_data = next(iter(vis_loader)).to(cfg.device)
15.     graph_embed, _ = model.embed_graph(graph_data.x, graph_data.edge_index, graph_data.batch)
16.     return graph_embed
```

*Listing 19. The use-case 1 implementation in our HW2VEC repository.*

The *preprocess* function on line 4 represents the HW2GRAPH's PRE_PROC module and is used specifically to locate the *entry point* top module in a hardware design *p*. It does this by flattening all the hardware codes into one file and removing all comments to search for the module that isn't instantiated by other modules. The module that is not instantiated by any others is renamed as the top module.

GRAPH_GEN and POST_PROC modules are used through HW2GRAPH's *process* function on line 5, which creates a graph generator to create a graph representation of the hardware design, either in DFG or AST format, and then transforms it into a NetworkX representation. For DFGs, we use PyVerilog modules like VerilogDataflowAnalyzer and VerilogGraphGenerator to retrieve the DFG before translating it into a NetworkX graph format. For ASTs, we parse the code into JSON-like structures which we then transform into a NetworkX graph.

The DATA_PROC module is instantiated in line 7 and we use the *process* function on line 8 to normalize the NetworkX graph and transform it into a PyTorch Geometric *Data* instance so we can perform graph-based ML operations. This is done by replacing all variable names with a high-level value type and calling the PyTorch Geometric *from_networkx* function to create an interfaceable *Data* instance with *X* and *A*.

We then initialize a pretrained GRAPH2VEC model on line 12 and retrieve the embedding of the graph through the *embed_graph* function on line 15. This function retrieves a graph embedding of the hardware design by taking a one-hot vector of the Data's features and performing a forward pass through the model's *GRAPH_CONV*, *GRAPH_POOL*, and *GRAPH_READOUT* layers to extract the final embedding, which can then be inspected.

## 3.2 Use-case 2: Hardware Trojan Detection

In this use-case, we demonstrate how to use HW2VEC to detect HT, which has been a major hardware security challenge for many years. An HT is an intentional, malicious modification of a circuit by an attacker [19]. The capability of detection at an early stage (particularly at RTL level) is crucial as removing

HTs at later stages could be very expensive. The majority of existing solutions rely on a golden HT-free reference or cannot generalize detection to previously unseen HTs. [10] proposes a GNN-based approach to model the circuit's behavior and identify the presence of HTs. In practice, we provide an implementation in `use_case_2.py` in our repository.

```
1.  if not cfg.data_pkl_path.exists():
2.      ''' converting graph using hw2graph '''
3.      nx_graphs = []
4.      hw2graph = HW2GRAPH(cfg)
5.      for hw_project_path in hw2graph.find_hw_project_folders():
6.          hw_graph = hw2graph.code2graph(hw_project_path)
7.          nx_graphs.append(hw_graph)
8.
9.      data_proc = DataProcessor(cfg)
10.     for hw_graph in nx_graphs:
11.         data_proc.process(hw_graph)
12.     data_proc.cache_graph_data(cfg.data_pkl_path)
13.
14. else:
15.     ''' reading graph data from cache '''
16.     data_proc = DataProcessor(cfg)
17.     data_proc.read_graph_data_from_cache(cfg.data_pkl_path)
```

*Listing 20. Initial data processing in use-case 2*

The code snippet above represents the data preparation and shows how we use HW2GRAPH to convert each hardware design *p* into a graph *g*. This code follows the same procedure as described in **use-case 1**, where each hardware design is transformed into a NetworkX graph representation in line 6 and then normalized and transformed into Data instances in line 11. If processed *Data* is already available, then it can be loaded immediately with line 17.

```
1.  TROJAN = 1
2.  NON_TROJAN = 0
3.
4.  all_graphs = data_proc.get_graphs()
5.  for data in all_graphs:
6.      if "TjFree" == data.hw_type:
7.          data.label = NON_TROJAN
8.      else:
9.          data.label = TROJAN
10.
11. train_graphs, test_graphs = data_proc.split_dataset(ratio=cfg.ratio, seed=cfg.seed,
        dataset=all_graphs)
12. train_loader = DataLoader(train_graphs, shuffle=True, batch_size=cfg.batch_size)
13. valid_loader = DataLoader(test_graphs, shuffle=True, batch_size=1)
```

*Listing 21. Data labelling and dataset splitting in use-case 2*

Then, in the dataset preparation shown above, we associate each *Data* instance with a label corresponding to whether a Trojan exists in the data in lines 4-9. Afterward, we split the entire dataset into two subsets for training and testing depending on user-defined parameters such as *ratio* and *seed*. These splits are transformed into *DataLoader* instances so that PyTorch Geometric utilities can be leveraged.

```
1.  model = GRAPH2VEC(cfg)
2.  if cfg.model_path != "":
```

```
3.        model_path = Path(cfg.model_path)
4.        if model_path.exists():
5.            model.load_model(str(model_path/"model.cfg"), str(model_path/"model.pth"))
6.    else:
7.        convs = [
8.            GRAPH_CONV("gcn", data_proc.num_node_labels, cfg.hidden),
9.            GRAPH_CONV("gcn", cfg.hidden, cfg.hidden)
10.       ]
11.       model.set_graph_conv(convs)
12.
13.       pool = GRAPH_POOL("sagpool", cfg.hidden, cfg.poolratio)
14.       model.set_graph_pool(pool)
15.
16.       readout = GRAPH_READOUT("max")
17.       model.set_graph_readout(readout)
18.
19.       output = nn.Linear(cfg.hidden, cfg.embed_dim)
20.       model.set_output_layer(output)
```

*Listing 22. Model configuration in use-case 2*

We then initialize the model and configure it as shown above. The user has the option to load a pre-trained model by altering the *model_path* argument in the command line, or they can define their own model's design and hyperparameters by directly modifying the code from lines 7-19. Users have the freedom to change out convolutional layers, pooling types, and readout types, among other hyperparameters such as hidden dimensions and pooling ratio.

```
1.    ''' training '''
2.    model.to(cfg.device)
3.    trainer = GraphTrainer(cfg, class_weights=data_proc.get_class_weights(train_graphs))
4.    trainer.build(model)
5.    trainer.train(train_loader, valid_loader)
6.
7.    ''' evaluating and inspecting '''
8.    trainer.evaluate(cfg.epochs, train_loader, valid_loader)
9.    vis_loader = DataLoader(all_graphs, shuffle=False, batch_size=1)
10.   trainer.visualize_embeddings(vis_loader, "./")
```

*Listing 23. Training and Evaluation pipeline in use-case 2*

**Use-case 2** then ends with the training and evaluating pipelines as shown in the code snippet above. We start training the model using the *GraphTrainer* class. Calling the *build* function in line 4 will assign the model to the trainer and create an Adam optimizer. Then, calling the *train* function in line 5 will train the graph embeddings with a cross-entropy loss function for the Adam optimizer while also performing mini-tests. Once the *GraphTrainer* has finished training, we can then call the *evaluate* function in line 8 and perform a final test to observe the performance of the model for TJ detection. To end, we call the *visualize_embeddings* function in line 10 to get the final graph embedding representation and metadata so that we can project these embeddings in Euclidean space and visualize what the model has learned.

### 3.3 Use-case 3: Hardware IP Piracy Detection

This use-case demonstrates how to leverage HW2VEC to confront another major hardware security challenge – determining whether one of the two hardware designs is stolen from the other or not. The IC supply chain has been so globalized that it exposes the IP providers to theft and illegal IP redistribution.

One state-of-the-art countermeasure embeds the signatures of IP owners on hardware designs (i.e., watermarking or fingerprinting), but it causes additional hardware overhead during manufacturing. Therefore, [9] addresses IP piracy by assessing the similarities between hardware designs with a GNN-based approach. Their approach models the behavior of a hardware design (in RTL or GLN) in graph representations. In practice, we provide the implementation in `use_case_3.py` in our repository.

The data preparation and model configuration aspects of **use-case 3** are the same as described for **use-case 2**. As such, this subsection will only cover the dataset preparation, training, and evaluation aspects of **use-case 3**.

```
1   SIMILAR = 1
2   DISSIMILAR = -1
3
4   data_proc.generate_pairs()
5   all_pairs = data_proc.get_pairs()
6   for pair_idx, pair in enumerate(all_pairs):
7       graph_a, graph_b = pair
8       if graph_a.hw_type == graph_b.hw_type:
9           all_pairs[pair_idx] += (SIMILAR,)
10      else:
11          all_pairs[pair_idx] += (DISSIMILAR,)
12
13  train_pairs, test_pairs = data_proc.split_dataset(cfg.ratio, cfg.seed, all_pairs)
14  train_loader = DataLoader(train_pairs, shuffle=True, batch_size=cfg.batch_size)
15  test_loader  = DataLoader(test_pairs, shuffle=True, batch_size=cfg.batch_size)
```

*Listing 24. Data labelling and dataset splitting in use-case 3*

In the dataset preparation for **use-case 3** shown above, we first create pairs of *Data* in line 4 using Python's built-in *combinations* function. Each pair instance is then associated with a label corresponding to whether the two Data instances in a pair are similar or not. Afterward, we split the entire dataset into two subsets for training and testing similar to how the dataset is split in **use-case 2.**

```
1   ''' training '''
2   model.to(cfg.device)
3   trainer = PairwiseGraphTrainer(cfg)
4   trainer.build(model)
5   trainer.train(train_loader, test_loader)
6
7   ''' evaluating and inspecting '''
8   trainer.evaluate(cfg.epochs, train_loader, test_loader)
9   vis_loader = DataLoader(data_proc.get_graphs(), shuffle=False, batch_size=1)
10  trainer.visualize_embeddings(vis_loader, "./")
```

*Listing 25. Training and Evaluation pipeline for use-case 3*

**Use-case 3** then ends with the training and evaluating pipelines as shown in the code snippet above. We start training the model using the *PairwiseGraphTrainer* class. The difference between this trainer and *GraphTrainer* in **use-case 2** is that *PairwiseGraphTrainer* utilizes a cosine similarity and cosine embedding loss instead of cross-entropy loss, and is trained in a pairwise manner. The training is otherwise the same as is described in **use-case 2.** Similarly, the evaluation pipeline follows the same procedures as described in **use-case 2.**

## 4. Experimental Results

In this section, we show our results of the various experiments using the use-case implementations described in section 3. We evaluated the results using an RTL dataset for HT detection (TJ-RTL) and both RTL and GLN datasets (IP-RTL and IP-GLN) for IP piracy detection. Sections 4.1 and 4.2 give more information about the datasets. Sections 4.3 and 4.4 go in depth about HT detection and IP piracy detection.

### 4.1 HW2VEC Evaluation: The TJ-RTL Dataset

We construct the TJ-RTL dataset by gathering the hardware designs with or without HT from the Trust-Hub.org benchmark [20]. From Trust-Hub, we collect three base circuits, AES, PIC, and RS232, and insert 34 varied types of HTs into them. We also include these HTs as standalone instances to the TJ-RTL dataset. Furthermore, we insert these standalone HTs into two other circuits (DES and RC5) and include the resulting circuits to expand the TJ-RTL dataset. Among the five base circuits, AES, DES, and RC5 are cryptographic cores that encrypt the input plaintext into the ciphertext based on a secret key. For these circuits, the inserted HTs can leak sensitive information (i.e., secret key) via side-channels such as power and RF radiation or degrade the performance of their host circuits by increasing the power consumption and draining the power supply. RS232 is an implementation of the UART communication channel, while the HT attacks on RS232 can affect the functionality of either transmitter or receiver or can interrupt/disable the communication between them. The PIC16F84 is a well-known Power Integrated Circuit (PIC) microcontroller, and the HTs for PIC fiddle with its functionality and manipulate the program counter register. Lastly, we create the graph datasets, DFG-TJ-RTL and AST-TJ-RTL, in which each graph instance is annotated with a TROJAN or NON_TROJAN label.

### 4.2 HW2VEC Evaluation: The IP-RTL and IP-GNL Datasets

To construct the datasets for evaluating piracy detection, we gather RTL and GLN of hardware designs in Verilog format. The RTL dataset includes common hardware designs such as single-cycle and pipeline implementation of MIPS processor which are derived from available open-source hardware design in the internet or designed by a group of in-house designers who are given the same specification to design a hardware in Verilog. The GLN dataset includes ISCAS'85 benchmark [21] instances derived from TrustHub. Obfuscation complicates the circuit and confuses reverse engineering but does not change the behavior of the circuit. Our collection comprises 50 distinct circuit designs and several hardware instances for each circuit design that sums up 143 GLN and 390 RTL codes. We form a graph-pair dataset of 19,094 similar pairs and 66,631 different pairs, dedicate 20% of these 85,725 pairs for testing and the rest for training. This dataset comprises pairs of hardware designs, labelled as PIRACY (positive) or NO-PIRACY (negative).

### 4.3 HW2VEC Evaluation: Hardware Trojan Detection

In this section, we present the results of our evaluation of the implementation in use-case 2. For performance metrics, we counted the True Positive (TP), False Negative (FN) and False Positive (FP) for deriving Precision P = TP/(TP + FP) and Recall R = TP/(TP + FN). R manifests the percentage of HT-infested samples that the model can identify. As the number of HT-free samples incorrectly classified as HT is also critical, we computed P, which indicates what percentage of the samples that model classifies as HT-infested actually contains HT. F1 score is the weighted average of precision and recall that better presents performance, calculated as F1 = 2 × P × R/(P + R).

To demonstrate whether the learned model can generalize the knowledge to handle the unknown or unseen circuits, we performed a variant leave-one-out cross-validation to experiment. We performed a train-test split on the TJ-RTL dataset by leaving one base circuit benchmark in the testing set and using the remaining circuits to train the model. We repeated this process for each base circuit and averaged the metrics we acquired from evaluating each testing set. The result is presented in Table 1, indicating that HW2VEC can reproduce comparable results to [10] in terms of F1 score (0.926 versus 0.940) if we use DFG as the graph representation. The difference in performance can be due to the use of different datasets. When using AST as the graph representation for detecting HT, HW2VEC performs worse in terms of F1 score, indicating that DFG is a better graph representation because it captures the data flow information instead of simply the syntactic information of a hardware design code. All in all, these results demonstrate that our HW2VEC can be leveraged for studying HT detection at design phases.

| Method | Graph | Dataset | Precision | Recall | F1 |
|--------|-------|---------|-----------|--------|-----|
| HW2VEC | DFG | RTL | 0.87334 | 0.98572 | 0.92596 |
| HW2VEC | AST | RTL | 0.90288 | 0.8 | 0.8453 |
| [10] | DFG | RTL | 0.923 | 0.966 | 0.940 |

*Table 1: The performance of HT detection using HW2VEC*

## 4.4 HW2VEC Evaluation: Hardware IP Piracy Detection

In this section, we discuss how we evaluated the power of HW2VEC in detecting IP piracy. We leveraged use-case 3, which examines the cosine-similarity score ŷ for each hardware design pair and produces the final prediction with the decision boundary. Using the IP-RTL dataset and the IP-GNL dataset, we generated graph-pair datasets by annotating the hardware designs that belong to the same hardware category as SIMILAR and the ones that belong to different categories as DISSIMILAR. We performed a train-test split on the dataset so that 80% of the pairs would be used to train the model. We computed the accuracy of detecting hardware IP piracy, which expresses the correctly predicted sample ratio and calculates the F1 score as the evaluating metrics. We refer to [9] for the selection of hyperparameters (stored in a YAML file).

The result is presented in Table 2, indicating that HW2VEC can reproduce comparable results to [9] in terms of piracy detection accuracy. When using DFG as the graph representation, HW2VEC underperforms [9] by 3% at RTL level and outperforms [9] by 4.2% at GLN level. Table 2 also shows a similar observation with Section 4.1 that using AST as the graph representation can lead to worse performance than using DFG. Figure 4 visualizes the graph embeddings that HW2VEC exports for every processed hardware design, allowing users to inspect the results manually. For example, by inspecting Figure 4, we may find a clear separation between mips_single_cycle and AES. Certainly, HW2VEC can perform better with more fine-tuning processes. However, the evaluation aims to demonstrate that HW2VEC can help practitioners study the problem of IP piracy at RTL and GLN levels.

| Method | Graph | Dataset | Accuracy | F1 |
|--------|-------|---------|----------|-----|

| HW2VEC | DFG | RTL | 0.9438 | 0.9277 |
|--------|-----|-----|--------|--------|
| HW2VEC | DFG | GLN | 0.9882 | 0.9652 |
| HW2VEC | AST | RTL | 0.9358 | 0.9183 |
| [9] | DFG | RTL | 0.9721 | - |
| [9] | DFG | GLN | 0.9461 | - |

*Table 2: The results of detecting IP piracy with HW2VEC*

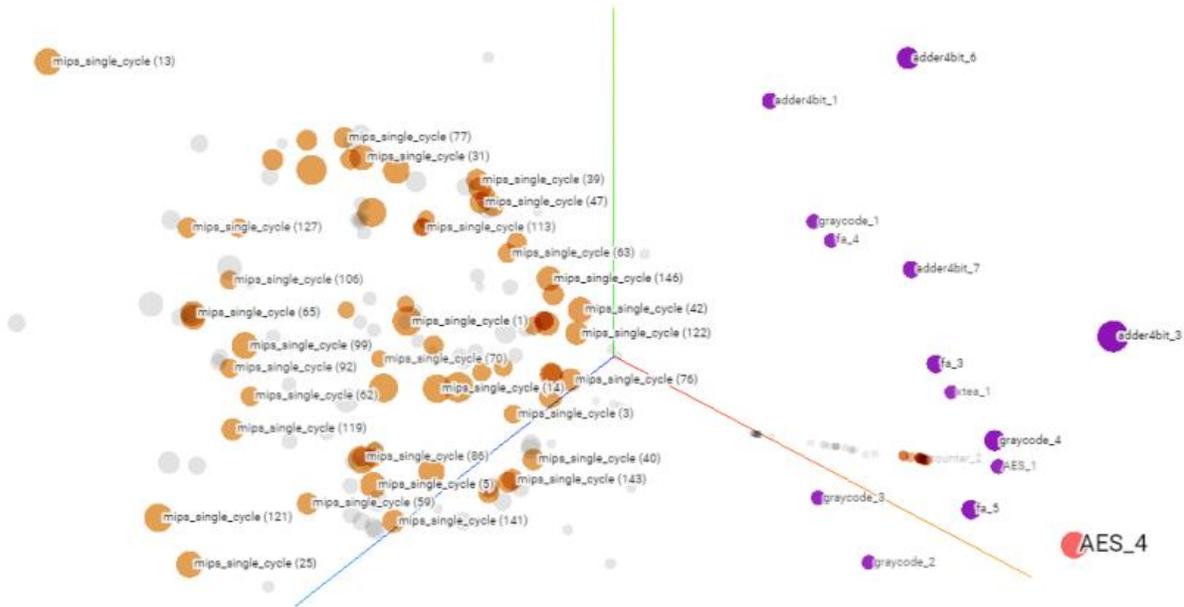

*Figure 4. The embedding visualization with 3D t-SNE*

## 5. Discussion & Conclusion

As technology continues to advance and grow more complex, the risk of cyberattacks will rise. To contribute to the hardware security research community, we propose HW2VEC, a flexible graph learning tool that provides an automated pipeline that allows for the extraction of graph representations from hardware design in either RTL or GLN. Our evaluation indicates that HW2VEC can be used to counteract Hardware Trojans and IP Piracy. We expect HW2VEC to provide easy access to apply graph learning approaches to hardware security applications for both practitioners and researchers.